\documentclass[10pt,letterpaper,english,aip, amsmath, amssymb, preprint]{revtex4-1}
\usepackage[T1]{fontenc}
\usepackage[latin9]{inputenc}
\usepackage{babel}
\usepackage{refstyle}
\usepackage{amsmath}
\usepackage{amssymb}
\usepackage{graphicx}
\usepackage{setspace}
\usepackage[unicode=true]
 {hyperref}

\makeatletter

\AtBeginDocument{\providecommand\enuref[1]{\ref{enu:#1}}}
\pdfpageheight\paperheight
\pdfpagewidth\paperwidth

\RS@ifundefined{subsecref}
  {\newref{subsec}{name = \RSsectxt}}
  {}
\RS@ifundefined{thmref}
  {\def\RSthmtxt{theorem~}\newref{thm}{name = \RSthmtxt}}
  {}
\RS@ifundefined{lemref}
  {\def\RSlemtxt{lemma~}\newref{lem}{name = \RSlemtxt}}
  {}

\usepackage{amsfonts}
\usepackage{amsmath,graphics, setspace}
\usepackage{braket}
\usepackage{hyperref}
\usepackage{color}
\usepackage[normalem]{ulem}
\usepackage[anythingbreaks]{breakurl}
\hypersetup{breaklinks=true}
\usepackage{graphicx}% Include figure files
\usepackage{dcolumn}% Align table columns on decimal point
\usepackage{bm}% bold math
\usepackage[T1]{fontenc}
\usepackage{mathptmx}
\allowdisplaybreaks
\hypersetup{colorlinks=true, linkcolor=blue}

\preprint{AIP/123-QED}

\makeatother

\begin{document}

\title{Removing instabilities in the hierarchical equations of motion: exact
and approximate projection approaches}

\author{Ian S. Dunn}

\author{Roel Tempelaar}

\author{David R. Reichman}
\email{drr2103@columbia.edu}

\selectlanguage{english}%

\affiliation{Department of Chemistry, Columbia University, 3000 Broadway, New
York, NY 10027, USA}

\date{\today}
\begin{abstract}
The hierarchical equations of motion (HEOM) provide a numerically
exact approach for computing the reduced dynamics of a quantum system
linearly coupled to a bath. We have found that HEOM contains temperature-dependent
instabilities that grow exponentially in time. In the case of continuous-bath
models, these instabilities may be delayed to later times by increasing
the hierarchy dimension; however, for systems coupled to discrete,
non-dispersive modes, increasing the hierarchy dimension does little
to alleviate the problem. We show that these instabilities can also
be removed completely at a potentially much lower cost via projection
onto the space of stable eigenmodes; furthermore, we find that for
discrete-bath models at zero temperature, the remaining projected
dynamics computed with few hierarchy levels are essentially identical
to the exact dynamics that otherwise might require an intractably
large number of hierarchy levels for convergence. Recognizing that
computation of the eigenmodes might be prohibitive, e.g. for large
or strongly-coupled models, we present a Prony filtration algorithm
that may be useful as an alternative for accomplishing this projection
when diagonalization is too costly. We present results demonstrating
the efficacy of HEOM projected via diagonalization and Prony filtration.
We also discuss issues associated with the nonnormality of HEOM.
\end{abstract}
\maketitle

\section{\label{sec:Introduction}Introduction}

A grand challenge in the physical sciences lies in modeling quantum
dynamics in the condensed phase\citep{School1998,May2011a,Weiss,Nitzan2006}.
Progress has often been made by using efficient approximate approaches
that invoke a \textquotedblleft system-bath\textquotedblright{} separation
and treat the bath classically\citep{Billing1975,Chen2016} or the
system-bath interactions perturbatively\citep{Redfield1965}. However,
in systems where quantum effects in the bath play a pronounced role
and where no small coupling or energy parameter may be identified,
exact solutions to a fully quantum system-bath model are desired.
One of the most successful computational methods for calculating exact
quantum dynamics is provided by the hierarchical equations of motion
(HEOM)\citep{Tanimura1989,Ishizaki2005,Ishizaki2009,Chen2009,Tanimura2014}.
First derived by Tanimura and Kubo\citep{Tanimura1989}, HEOM is a
reformulation of the Feynman-Vernon influence functional approach
to quantum dissipative dynamics\citep{Feynman1963,Tanimura1989,Makri1995,Makri1995a,Ishizaki2005,Ishizaki2009,Liu2014,Chen2015},
and its solution yields the exact reduced dynamics of linearly coupled
system-bath models. HEOM has successfully addressed a variety of applications;
yet, its scope is limited since the original formulation of HEOM requires
the bath to be represented by a continuous spectral density. Such
a constraint is often inappropriate for describing phenomena captured
in venerable models of quasi-particle dynamics in organic molecular
crystals\citep{Spano2010,Bakulin2016,Tempelaar2018,Fujihashi2017,Morrison2017},
coupled excitonic and vibrational motion in light harvesting complexes\citep{Womick2011,Christensson2012,Tiwari2013,Tempelaar2014},
and transport in polar crystals with narrow phonon bandwidths\citep{Frohlich1954,Feynman1955,Feynman1962,Devreese2009}.
In these cases where discrete bath modes play a significant role,
efficient and exact methods for computing quantum dynamics are desirable.

Motivated by the aforementioned computational challenges, several
groups have recently explored novel formulations of HEOM that treat
a discrete spectral density\citep{Liu2014,Chen2015}. There are several
potential advantages in developing a \textquotedblleft discrete-bath
HEOM\textquotedblright . First, even in the discrete-bath formulation,
HEOM retains the benefit that the reduced equations automatically
include all possible bath excitations. As a result, HEOM eliminates
the issue of basis-set convergence present in techniques such as exact
diagonalization and matrix product states, and instead relies on convergence
with respect to hierarchy depth (number of hierarchy levels, see \ref{subsec:The-hierarchical-equations}).
Second, including discrete bath modes in HEOM serves to expand the
ability of this powerful methodology to tackle an important class
of problems, with the added benefit that several popular HEOM software
packages such as \textit{Parallel Hierarchy Integrator\citep{Strumpfer2012}}
(\textit{PHI}), \textit{pyrho}\citep{Berkelbach}, and potentially
also \textit{GPU-HEOM\citep{Kreisbeck2011}} can be readily adapted
for the discrete-bath case.
\begin{figure}[h]
\includegraphics{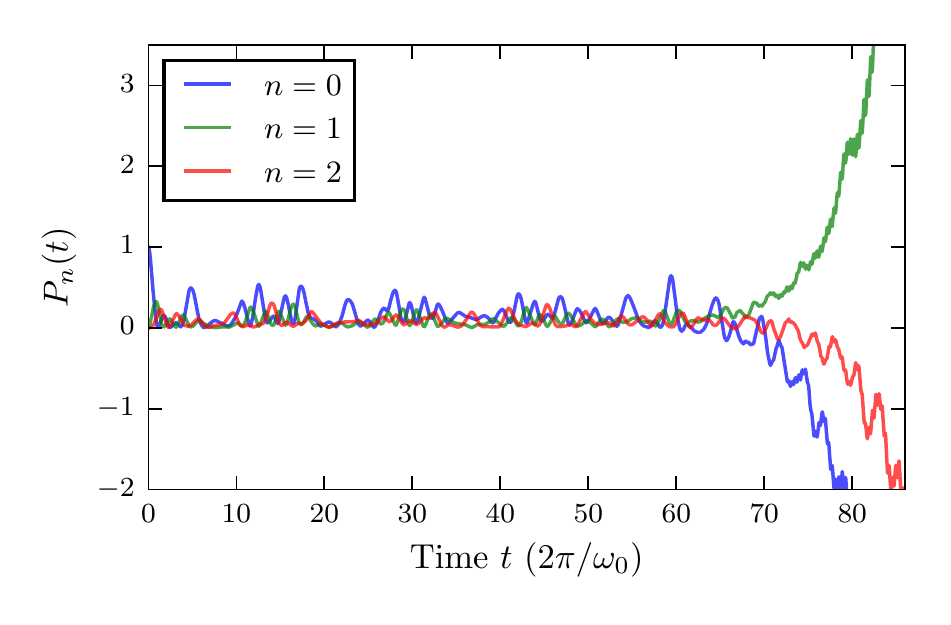}\caption{Time evolution of electron density for a 10-site periodic Holstein
model with a dispersive phonon band (as defined in eqs. 1-4 of Chen
et al.\citep{Chen2015}) calculated via HEOM with 5 hierarchy levels.
$P_{n}(t)\equiv\langle a_{n}^{\dagger}(t)a_{n}(t)\rangle$. Model
parameters: $g=0.1,\omega_{0}=1,W=0.5,J=0.2,T=0$. This is an extension
to longer times of the simulation performed in Fig. 1a of Chen et
al., which was originally run to time $12\pi/\omega_{0}$\citep{Chen2015}.
Calculations were performed using a modified version of \textit{PHI\citep{Strumpfer2012}.}}
\label{Fig. 1}
\end{figure}

Nonetheless, existing formulations of discrete-bath HEOM are not without
serious problems that we will explore and partially remedy in this
work. Recently, Chen et al.\citep{Chen2015} published discrete-bath
HEOM simulations which converge with few hierarchy levels to the exact
reduced dynamics for a 10-site Holstein model. We find that upon extending
the time of these simulations, the dynamics are eventually plagued
by the abrupt onset of an exponential instability. This instability
is illustrated in Fig. \ref{Fig. 1}. Unfortunately, increasing the
number of hierarchy levels does little to delay this instability to
later times and comes with a great computational cost. In the interest
of honing HEOM to be a useful tool for modeling the dynamics of discrete-bath
models, in this work we explore such instabilities and discuss approaches
for removing them to facilitate longer-time simulations without the
need for a large hierarchy depth. We also show that similar instabilities
exist in the original \textquotedblleft continuous-bath HEOM\textquotedblright{}
at low temperatures. Our main finding, which is explicit in the discrete-bath
case and plausible in the continuous-bath case, is that one can remove
these instabilities \textit{without altering the exact long-time dynamics}.

In this work we proceed as follows. In \ref{sec:Theory} we introduce
the condensed phase models which we study, as well as the HEOM that
describe their dynamics. In \ref{sec:Spectral-analysis} we introduce
our spectral approach for studying the stability of HEOM and illustrate
several stable and unstable examples. In \ref{sec:Temperature-Dependence-of},
we then apply our stability analysis to both discrete-bath and continuous-bath
HEOM at nonzero temperatures, providing a brief discussion of instabilities
in low-temperature continuous-bath HEOM that has only been alluded
to\citep{Moix2013} in the vast HEOM literature. In \ref{sec:Projecting-away-instabilities}
we discuss a diagonalization approach for exactly projecting out the
instabilities in discrete-bath HEOM. In \ref{sec:Projecting-away-instabilities-1}
we present an iterative method for accomplishing this projection that
does not require the full diagonalization of the HEOM. Finally, in
\ref{sec:Nonnormality-of-HEOM} we comment on some difficulties that
may arise due to the nonnormality of HEOM.
\begin{spacing}{1.2}

\section{Theory\label{sec:Theory}}
\end{spacing}

\subsection{Model Hamiltonians}

To demonstrate instabilities in HEOM we consider several standard
system-bath models for open quantum dynamics. Each of these models
takes the form
\begin{align}
\hat{H} & =\hat{H}_{s}+\hat{H}_{b}+\hat{H}_{sb},
\end{align}
where $\hat{H}_{s}$ describes the degrees of freedom (DOFs) of the
reduced system, $\hat{H}_{b}$ describes the DOFs of the phonon bath,
and $\hat{H}_{sb}$ describes the coupling between the system and
bath DOFs. In following sections, we will refer to the delta-function
spin-boson model (DSB), the Holstein model\citep{Holstein1959,Holstein1959a},
and the Su-Schrieffer-Heeger (SSH) model\citep{Su1979} as discrete-bath
models. Likewise, we will refer to the continuous spin-boson model
(CSB) as a continuous-bath model. $\hat{a}$ and $\hat{b}$ are electron
and phonon annihilation operators, respectively. We work in dimensionless
units and set $\hbar=1$.

\subsubsection{Spin-Boson Model}

In this work, we consider a spin-boson model\citep{Weiss,Chen2016}
with no energy bias between sites. The system has two energy levels,
\begin{align}
\hat{H}_{s} & =-J\left(\hat{a}_{1}^{\dagger}\hat{a}_{0}+\hat{a}_{0}^{\dagger}\hat{a}_{1}\right),
\end{align}
where $J$ is the inter-site coupling constant. This two-level system
interacts with a harmonic oscillator bath
\begin{align}
\hat{H}_{b} & =\sum_{j=1}^{N_{b}}\omega_{j}\hat{b}_{j}^{\dagger}\hat{b}_{j},
\end{align}
where $\omega_{j}$ is the frequency of the $j$-th bath mode. We
consider two forms of this model. In the DSB there is only one bath
oscillator, such that $N_{b}=1$, $\omega_{j}\equiv\omega_{0}$, and
$b_{j}\equiv b$. For the DSB, the coupling is defined as
\begin{align}
\hat{H}_{sb} & =-g\omega_{0}\hat{V}\left(\hat{b}+\hat{b}^{\dagger}\right),
\end{align}
where
\begin{align}
\hat{V} & =\hat{a}_{0}^{\dagger}\hat{a}_{0}-\hat{a}_{1}^{\dagger}\hat{a}_{1}.
\end{align}
In the CSB an infinite number of bath oscillators are included. For
the CSB, the coupling is defined as
\begin{align}
\hat{H}_{sb} & =-\hat{V}\sum_{j=1}^{N_{b}}c_{j}\left(\hat{b}_{j}+\hat{b}_{j}^{\dagger}\right),
\end{align}
where
\begin{align}
\hat{V} & =\hat{a}_{0}^{\dagger}\hat{a}_{0}-\hat{a}_{1}^{\dagger}\hat{a}_{1}.
\end{align}
The coefficients $c_{j}$ are fixed via the spectral density
\begin{align}
J(\omega) & =\sum_{j=1}^{N_{b}}c_{j}^{2}\delta\left(\omega-\omega_{j}\right),
\end{align}
which we choose in this work to be the Debye spectral density
\begin{align}
J(\omega) & =\frac{2}{\pi}\frac{\lambda\gamma\omega}{\omega^{2}+\gamma^{2}}.
\end{align}
In practice, the spin-boson model may be used as a coarse-grained
description for any system that is well approximated by a two-level
system coupled linearly to a harmonic bath. Applications of the spin-boson
model are enumerated by Weiss and include the description of qubits,
tunneling phenomena, and electron transfer processes\citep{Weiss}.

\subsubsection{Holstein model}

We also consider a one-dimensional Holstein model\citep{Holstein1959,Holstein1959a,Mahan2000}
with periodic boundary conditions. The system is described by a tight-binding
Hamiltonian
\begin{align}
\hat{H}_{s} & =-J\sum_{n=1}^{N}\hat{a}_{n}^{\dagger}\left(\hat{a}_{n+1}+\hat{a}_{n-1}\right),
\end{align}
interacting with a harmonic oscillator bath 
\begin{align}
\hat{H}_{b} & =\omega_{0}\sum_{n=1}^{N}\hat{b}_{n}^{\dagger}\hat{b}_{n},
\end{align}
with site-diagonal coupling
\begin{align}
\hat{H}_{sb} & =-g\omega_{0}\sum_{n=1}^{N}\hat{V}_{n}\left(\hat{b}_{n}+\hat{b}_{n}^{\dagger}\right),\label{eq:-6}
\end{align}
where
\begin{align}
\hat{V}_{n} & =\hat{a}_{n}^{\dagger}\hat{a}_{n}.
\end{align}
The Holstein model, termed a ``molecular crystal model,'' was introduced
to extend the conventional continuum treatment of polarons\citep{Pekar1946,Frohlich1954}
to account for the deformation of a discrete lattice\citep{Holstein1959,Holstein1959a}.
It reflects the decoupled nature of sites in a molecular crystal by
including only local electron-phonon coupling under the assumption
of Einstein phonons. The Holstein model has the advantage that it
can be used for a range of coupling strengths to describe large and
small polarons, alike\citep{Holstein1959,Holstein1959a}. For an excellent
review that discusses the relation between the Holstein model and
the Fröhlich model as well as the DSB, see Devreese and Alexandrov\citep{Devreese2009}.
In addition to modeling electron dynamics, the Holstein model has
also seen great success in modeling Frenkel exciton dynamics in organic
molecular crystals\citep{Merrifield1964,Spano2010,Hestand2018}. 

\subsubsection{Su-Schrieffer-Heeger model (SSH)}

As an alternative to the Holstein model, we also briefly consider
the SSH model\citep{Su1979} with periodic boundary conditions, which
differs from the Holstein model in its off-diagonal system-bath coupling
\begin{align}
\hat{H}_{sb} & =-gJ\sum_{n=1}^{N}\hat{V}_{n}\left[\left(\hat{b}_{n}+\hat{b}_{n}^{\dagger}\right)-\left(\hat{b}_{n+1}+\hat{b}_{n+1}^{\dagger}\right)\right],\label{eq:-7}
\end{align}
where
\begin{align}
\hat{V}_{n} & =\hat{a}_{n}^{\dagger}\hat{a}_{n+1}+\hat{a}_{n+1}^{\dagger}\hat{a}_{n}.
\end{align}
The SSH model was originally proposed to describe solitons in polyacetylene\citep{Su1979},
and has also been employed to model charge transport in crystalline
organic semiconductors\citep{DeFilippis2015}. The SSH coupling (\ref{eq:-7})
accounts for the modulation of electron-hopping rates based on the
variable nuclear distance between sites. When the SSH coupling (\ref{eq:-7})
is combined with the Holstein coupling (\ref{eq:-6}), the resulting
model which includes both local deformation and phonon-mediated hopping
is known as the Holstein-Peierls model\citep{Munn1985}.

\subsection{The hierarchical equations of motion\label{subsec:The-hierarchical-equations}}

We now formally define HEOM, the exact quantum dynamics method of
interest in this work. HEOM consists of a set of coupled linear differential
equations that govern the time evolution of a hierarchy of indexed
matrices. At the root of the hierarchy lies the reduced density matrix
of the system, 
\begin{align}
\hat{\sigma}(t) & =\hat{\rho}_{0,..,0}(t).
\end{align}
The dynamics of the DSB, Holstein model, and SSH model are described
by the following HEOM\citep{Liu2014,Chen2015}.
\begin{align}
\frac{d}{dt}\hat{\rho}_{m_{1\pm},...,m_{N_{b}\pm}}(t) & =-i\mathcal{L}\hat{\rho}_{m_{1\pm},...,m_{N_{b}\pm}}(t)\nonumber \\
 & -i\sum_{n=1}^{N_{b}}\omega_{0}\left(m_{n-}-m_{n+}\right)\hat{\rho}_{m_{1\pm},...,m_{N_{b}\pm}}(t)\nonumber \\
 & +\sum_{n=1}^{N_{b}}\biggr[\Phi_{n}\biggr(\hat{\rho}_{m_{1\pm},...,m_{n+}+1,...,m_{N_{b}\pm}}(t)\nonumber \\
 & +\hat{\rho}_{m_{1\pm},...,m_{n-}+1,...,m_{N_{b}\pm}}(t)\biggr)\nonumber \\
 & +m_{n+}\Theta_{n+}\hat{\rho}_{m_{1\pm},...,m_{n+}-1,...,m_{N_{b}\pm}}(t)\nonumber \\
 & +m_{n-}\Theta_{n-}\hat{\rho}_{m_{1\pm},...,m_{n-}-1,...,m_{N_{b}\pm}}(t)\biggr],\label{eq:}
\end{align}
where
\begin{align}
\mathcal{L} & =[\hat{H}_{s},...],
\end{align}
and
\begin{align}
\Phi_{n} & =[\hat{V}_{n},...].
\end{align}
For the Holstein model and the DSB,
\begin{align}
\Theta_{n\pm} & =-\frac{\left(g\omega_{0}\right)^{2}}{2}\left([\hat{V}_{n},...]\coth\left(\frac{\beta\omega_{0}}{2}\right)\mp\{\hat{V}_{n},...\}\right),
\end{align}
while for the SSH model
\begin{align}
\Theta_{n\pm} & =-\frac{\left(gJ\right)^{2}}{2}\biggr([\hat{V}_{n}-\hat{V}_{n-1},...]\coth\left(\frac{\beta\omega_{0}}{2}\right)\nonumber \\
 & \mp\{\hat{V}_{n}-\hat{V}_{n-1},...\}\biggr).
\end{align}
For the Holstein and SSH models, $N_{b}=N$. For the DSB, $N_{b}=1$
and $\hat{V}_{n}=\hat{V}$. Throughout this paper we define the inverse
temperature $\beta=\left(k_{B}T\right)^{-1}$ and work in units where
$k_{B}=1$.

The $l$-th hierarchy level consists of all matrices $\hat{\rho}_{m_{1\pm},...,m_{N_{b}\pm}}(t)$
in Eq. (\ref{eq:}) for which 
\begin{align}
\sum_{n+=1}^{N_{b}}m_{n+}+\sum_{n-=1}^{N_{b}}m_{n-} & =l.
\end{align}
In this study we truncate this infinite hierarchy of coupled differential
equations after $L$ hierarchy levels with a ``time-nonlocal'' closure\citep{Chen2009,Liu2014,Chen2015},
where we set 
\begin{align}
\hat{\rho}_{m_{1\pm},...,m_{N_{b}\pm}}(t) & =0
\end{align}
 for 
\begin{align}
\sum_{n+=1}^{N_{b}}m_{n+}+\sum_{n-=1}^{N_{b}}m_{n-} & \geq L.\label{eq:-11}
\end{align}
In practice, solutions to (\ref{eq:}) are to be converged with respect
to the hierarchy depth $L$. 

The CSB is described by the following HEOM\citep{Ishizaki2005}, which
are similar in structure to Eq. (\ref{eq:}) but produce markedly
different dynamics due to the incorporation of an infinite bath.
\begin{align}
\frac{d}{dt}\hat{\rho}_{m_{0},...,m_{K}}(t) & =\left[-i\mathcal{L}-\sum_{k=0}^{K}m_{k}\nu_{k}\right]\hat{\rho}_{m_{0},...,m_{K}}(t)\nonumber \\
 & +\sum_{k=0}^{K}\biggr[\Phi\hat{\rho}_{m_{0},...,m_{k}+1,...,m_{K}}(t)\nonumber \\
 & +\sum_{k=0}^{K}m_{k}\Theta_{k}\hat{\rho}_{m_{0},...,m_{k}-1,...,m_{K}}(t)\biggr],\label{eq:-1}
\end{align}
where
\begin{align}
\mathcal{L} & =[\hat{H}_{s},...],\\
\Phi & =[\hat{V},...],\\
\Theta_{0} & =-\lambda\gamma\left(\cot\left(\frac{\beta\gamma}{2}\right)[\hat{V},...]-i\{\hat{V},...\}\right),\label{eq:-9}\\
\Theta_{k\neq0} & =-\frac{4\lambda\gamma\nu_{k}}{\beta}\frac{1}{\nu_{k}^{2}-\gamma^{2}}[\hat{V},...],\\
\nu_{0} & =\gamma,
\end{align}
and
\begin{align}
\nu_{k\neq0} & =\frac{2\pi k}{\beta}.
\end{align}

Eq. (\ref{eq:-1}) incorporates an infinite Matsubara series, resulting
from a high-temperature expansion, that is closed via truncation after
$K$ Matsubara terms. Another popular closure for the Matsubara series
in HEOM was derived by Ishizaki and Tanimura\citep{Ishizaki2005}.
The Ishizaki-Tanimura closure approximately accounts for Matsubara
terms with $k>K$ (for sufficiently large $K$) by replacing rapidly
decaying factors of $\nu_{k}e^{-\nu_{k}t}$ with $\delta\left(t\right)$;
this closure also has the added benefit of improved stability, although
instabilities are still present. However, in the interest of using
a continuous-bath HEOM that closely resembles the discrete-bath formulation
in Eq. (\ref{eq:}) we will not employ the Ishizaki-Tanimura closure
in this work.

Here, the $l$-th hierarchy level consists of all matrices $\hat{\rho}_{m_{0},...,m_{K}}(t)$
in Eq. (\ref{eq:}) for which 
\begin{align}
\sum_{k=0}^{K}m_{k} & =l.
\end{align}
Again, we use a time-nonlocal closure after $L$ hierarchy levels\citep{Ishizaki2005,Chen2009}
such that 
\begin{align}
\hat{\rho}_{m_{0},...,m_{K}}(t) & =0
\end{align}
 for 
\begin{align}
\sum_{k=0}^{K}m_{k} & \geq L.\label{eq:-12}
\end{align}
In practice, solutions to (\ref{eq:-1}) are to be converged with
respect to the hierarchy depth $L$, as well as the number of Matsubara
terms $K$. 

Alternate closures exist for Eqs. (\ref{eq:}) and (\ref{eq:-1})
besides those shown in Eqs. (\ref{eq:-11}) and (\ref{eq:-12}). Continuous-bath
HEOM studies regularly employ a closure that relies on the exponential
suppression of deeper hierarchy levels\citep{Tanimura1991,Ishizaki2005};
we do not investigate this closure here since it is not applicable
to discrete-bath models. Furthermore, the time-local closure\citep{Xu2005,Chen2009}
is applicable for both discrete-bath HEOM and continuous-bath HEOM;
however, since it does not appear to suppress instabilities in discrete-bath
HEOM and also is not amenable to the spectral analysis in \ref{sec:Spectral-analysis},
we do not explore the time-local closure here.

For the initial condition employed in this work, similar to many other
studies, we set the population of the first site $\rho_{0,...,0}^{00}(t=0)=1$,
and all the other hierarchical matrix elements are set to zero.
\begin{spacing}{1.2}

\section{Spectral analysis\label{sec:Spectral-analysis}}
\end{spacing}

The HEOM presented in Eqs. (\ref{eq:}) and (\ref{eq:-1}) may be
represented as linear systems of the form
\begin{align}
\frac{d}{dt}\vec{\rho}(t) & =A\vec{\rho}(t),\label{eq:-2}
\end{align}
where $\vec{\rho}(t)$ is a vector containing all the hierarchy elements
$\rho_{m_{1\pm},...,m_{N_{b}\pm}}^{ij}(t)$ or $\rho_{m_{0},...,m_{K}}^{ij}(t)$
and $A$ is a nonnormal matrix containing the coupling between all
of the hierarchy elements. With this flattened representation of HEOM,
spectral analysis can be used to study the stability of solutions.
The solution to Eq. (\ref{eq:-2}) may be written as
\begin{align}
\vec{\rho}(t) & =e^{At}\vec{\rho}(0).\label{eq:-4}
\end{align}
Assuming $A$ to be a diagonalizable matrix and employing the eigen-decomposition
of $A$, we can write Eq. (\ref{eq:-4}) as 
\begin{align}
\vec{\rho}(t) & =Ve^{\Lambda t}V^{-1}\vec{\rho}(0),\label{eq:-5}
\end{align}
where $\Lambda$ is a diagonal matrix containing the eigenvalues $\{\lambda_{i}\}$
of $A$ and $V$ is a matrix whose columns $\vec{v}_{i}$ are the
normalized eigenvectors of $A$. We can write Eq. (\ref{eq:-5}) in
the eigenbasis of $A$ as
\begin{align}
\vec{\rho}(t) & =\sum_{i}d_{i}e^{\lambda_{i}t}\vec{v}_{i},
\end{align}
where the components of $\vec{d}=V^{-1}\vec{\rho}(0)$ are the expansion
coefficients of $\vec{\rho}(0)$ in the eigenbasis of $A$. From this
representation, it is clear that terms in the sum with $\Re[\lambda_{i}]>0$
are asymptotically unstable, and will be referred to here as the unstable
modes.
\begin{figure}[h]
\includegraphics{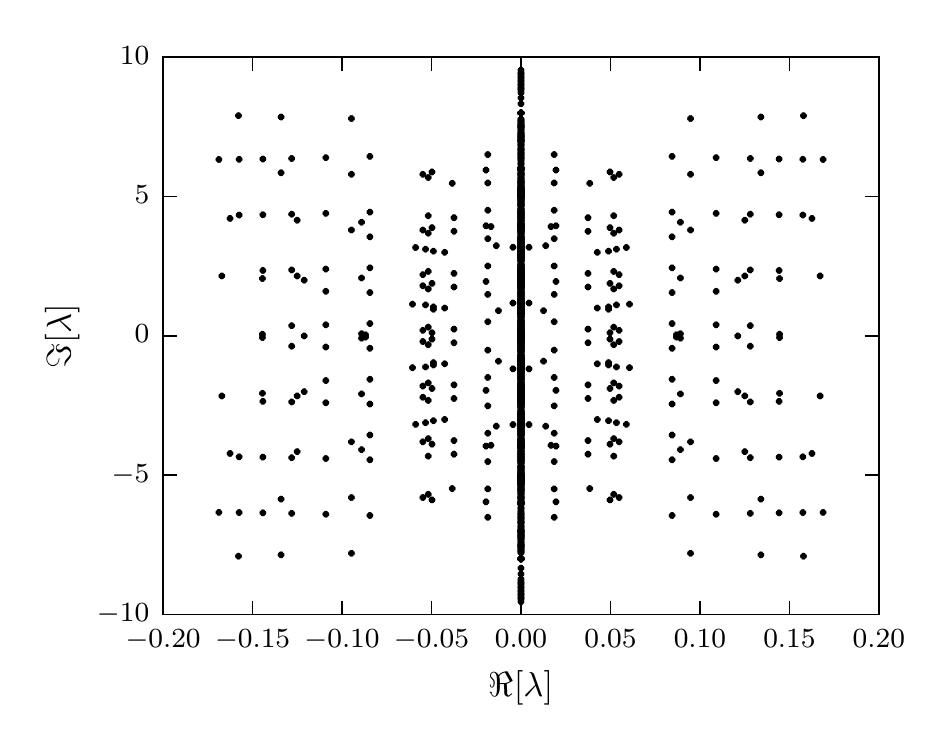}\caption{Spectrum of $A$ for a 2-site Holstein model at $T=0$ with $L=9$.
Model parameters: $g=0.5,\omega_{0}=1,J=0.2$.}

\label{Fig. 3}
\end{figure}
\\

\subsection{Unstable HEOM}

\begin{figure}[h]
\includegraphics{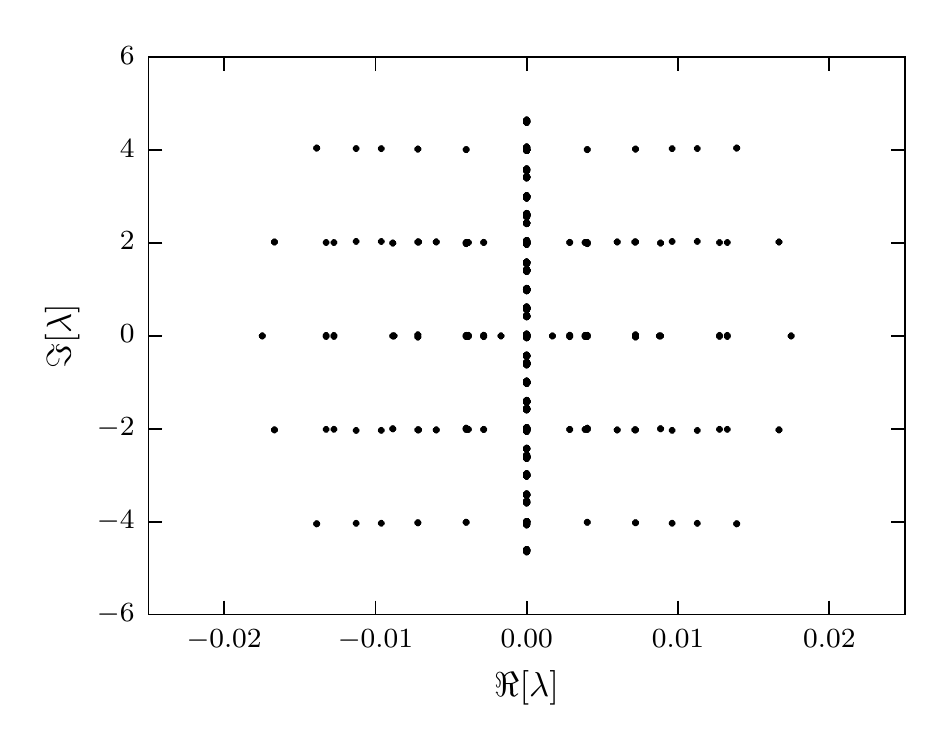}\caption{Spectrum of $A$ for a 3-site SSH model at $T=0$ with $L=5$. Model
parameters: $g=0.5,\omega_{0}=1,J=0.2$.}

\label{Fig. 4}
\end{figure}
 
\begin{figure}[h]
\includegraphics{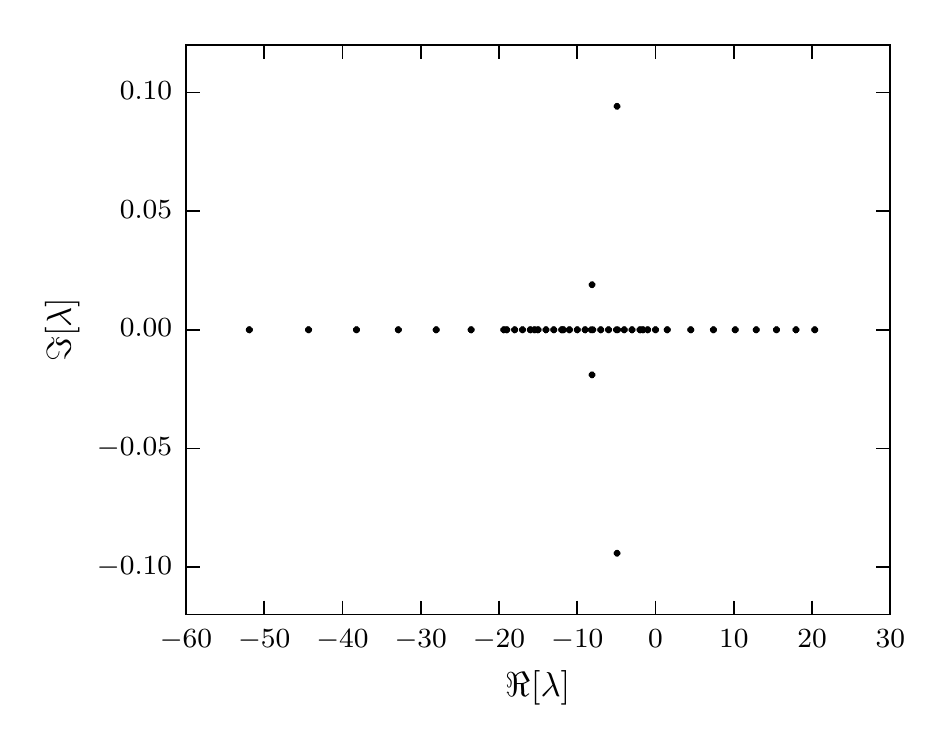}\caption{Spectrum of $A$ for the CSB with $L=20,K=0$. Model parameters: $\sqrt{\lambda}=0.3,\gamma=1,J=0.2,T=0.16$.}

\label{Fig. 16}
\end{figure}
In Fig. \ref{Fig. 3} and Fig. \ref{Fig. 4} we calculate the eigen-decomposition
for the flattened representation of Eq. (\ref{eq:}) and plot $\{\lambda_{i}\}$
for two discrete-bath models: the 2-site Holstein model and the 3-site
SSH model. Likewise, in Fig. \ref{Fig. 16} we do the same for the
CSB described by Eq. (\ref{eq:-1}). Notice that eigenvalues are present
in the right half-plane in all three cases: these eigenvalues correspond
to unstable modes. Furthermore, computing $\vec{d}$, the decomposition
of $\vec{\rho}(0)$ in the eigenbasis of $A$, reveals that some of
these unstable modes have nonzero weights in the initial condition,
leading to asymptotic instability in the dynamics. In the following
sections we will interpret these unstable modes and discuss computational
strategies for removing them.

While we have shown in Fig. \ref{Fig. 16} a particularly unstable
example of continuous-bath HEOM, in many practical continuous-bath
cases one can suppress any instabilities by converging with respect
to $L$ and $K$ and employing the Ishizaki-Tanimura closure\citep{Ishizaki2005}
for the Matsubara series. \textit{To the contrary, instabilities are
much harder to suppress via convergence with respect to $L$ in discrete-bath
HEOM.}

\subsection{Asymptotically Stable HEOM}

\begin{figure}[t]
\includegraphics{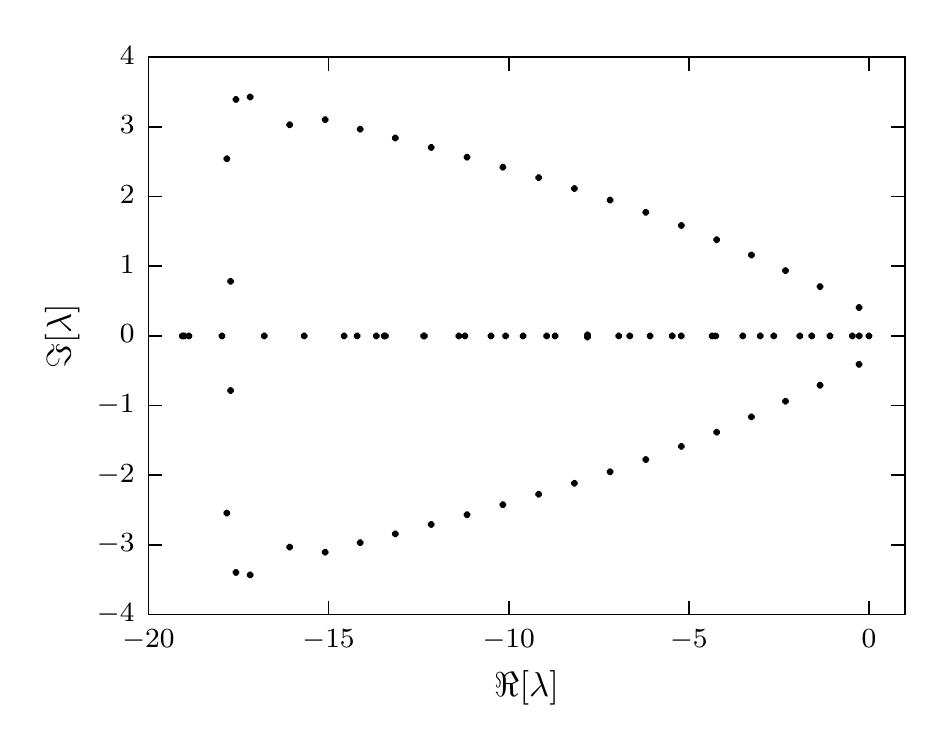}\caption{Spectrum of $A$ for the CSB with $L=20,K=0.$ Model parameters: $\sqrt{\lambda}=0.5,\gamma=1,J=0.2,T=0.4$.}

\label{Fig. 2}
\end{figure}

To show an example of asymptotically stable HEOM, we will now contrast
the former unstable examples with the HEOM for the CSB (\ref{eq:-1})
at a temperature where no unstable modes are present. In Fig. \ref{Fig. 2}
we again show the spectrum of the flattened representation of Eq.
(\ref{eq:-1}) 
(the HEOM for the CSB), only this time at a higher temperature. Since
the high-temperature $K=0$ approximation for continuous-bath HEOM
is known to be equivalent to the Zusman equation\citep{Shi2009},
we note the resemblance here to the eigentree structure from the Zusman
equation spectral analysis reported by Jung et al.\citep{Jung1999}.
It is clear that all eigenvalues are confined to the left half-plane
of Fig. \ref{Fig. 2}. As a result, the corresponding dynamics are
asymptotically stable.

At this point, we would be remiss to not acknowledge the structure
and symmetry present in the spectral plots of Figs. \ref{Fig. 3},
\ref{Fig. 4}, \ref{Fig. 16}, and \ref{Fig. 2}. While we omit a
discussion of the spectral dependence on hierarchy depth, number of
sites, and choice of model, we exemplify these dependences in the
animations shown in the \href{http://www.columbia.edu/cu/chemistry/groups/reichman/heomstability.html}{supplementary material}.

\section{Temperature Dependence of HEOM Spectra\label{sec:Temperature-Dependence-of}}

\begin{figure*}
\includegraphics{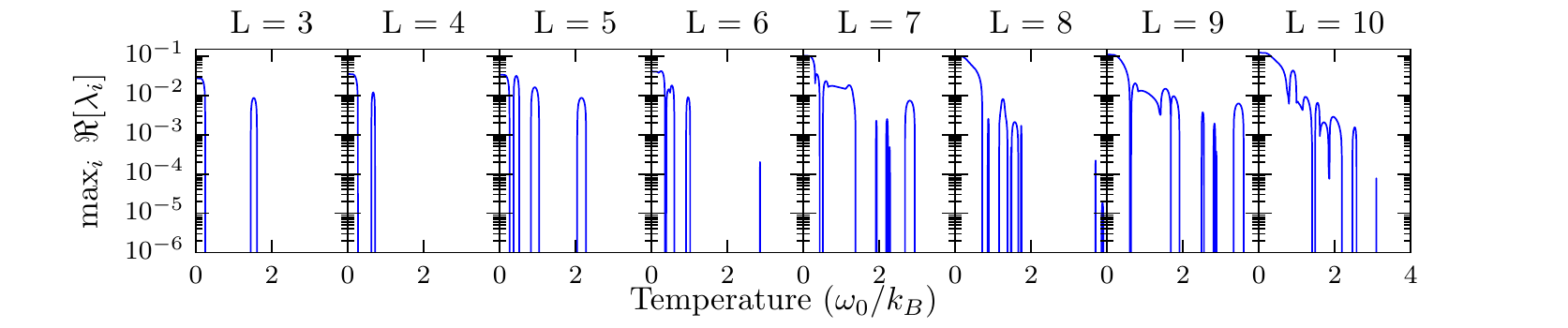}\caption{Real part of most unstable eigenvalue for the DSB, plotted as a function
of temperature for a range of hierarchy depths. Model parameters:
$g=0.4,\omega_{0}=1,J=0.2$.}
\label{Fig. 13}
\end{figure*}

\begin{figure}[h]
\includegraphics{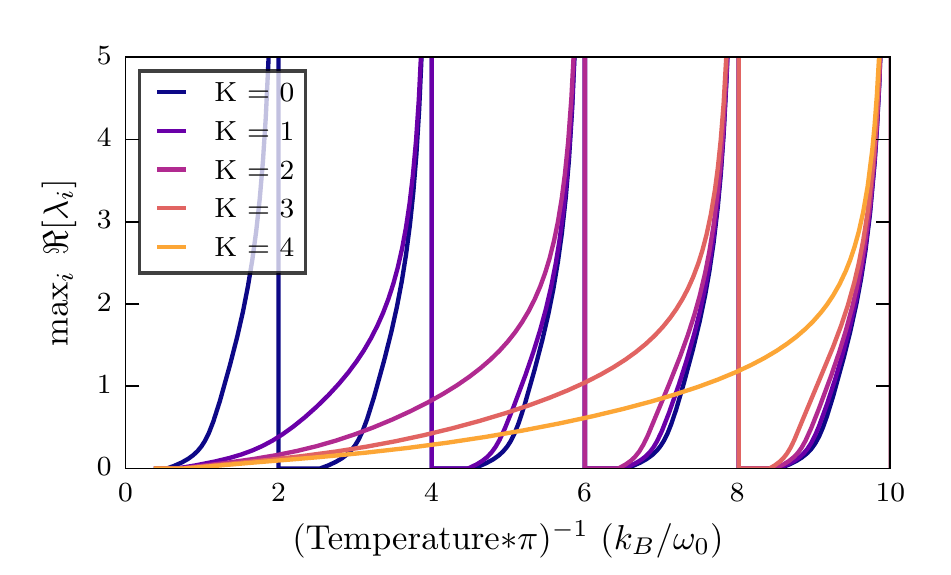}\caption{Real part of most unstable eigenvalue for the CSB, plotted as a function
of temperature for a variety of Matsubara dimensions. Model parameters:
$\sqrt{\lambda}=0.8,\gamma=1,J=0.2$. $L=3$.}
\label{Fig. 14}

\end{figure}

\begin{figure}[h]
\includegraphics{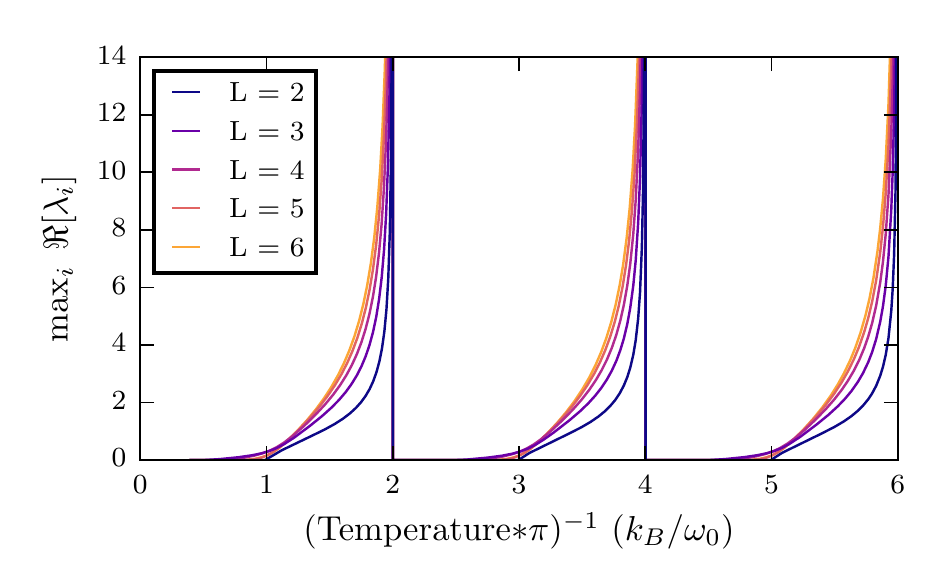}\caption{Real part of most unstable eigenvalue for the CSB, plotted as a function
of temperature for a variety of hierarchy depths. Model parameters:
$\sqrt{\lambda}=0.8,\gamma=1,J=0.2$. $K=0$.}

\label{Fig. 15}

\end{figure}

The spectra of both discrete-bath HEOM and continuous-bath HEOM admit
a rich temperature dependence. In Figs. \ref{Fig. 13}, \ref{Fig. 14},
and \ref{Fig. 15} 
we plot for the DSB and the CSB the real part of the most unstable
eigenvalue, $\max_{i}\Re\left[\lambda_{i}\right]$, as a function
of temperature. Note the similarity between the qualitative behavior
in Fig. \ref{Fig. 13} and Figs. \ref{Fig. 14} and \ref{Fig. 15}.
Both discrete-bath and continuous-bath HEOM reveal unstable regions,
i.e. temperature ranges where $\max_{i}\Re\left[\lambda_{i}\right]>0$,
intermingled with stable regions. Furthermore, both have unstable
regions concentrated at lower temperatures. 

While there are gross similarities between these cases, there are
several important differences of note. First, the behavior shown in
Fig. \ref{Fig. 13} seems to be piecewise continuous while that of
Figs. \ref{Fig. 14} and \ref{Fig. 15} contains many asymptotes.
A simple analytically solvable example that helps us rationalize the
appearance of these asymptotes for the CSB will be discussed in the
\hyperref[Appendix]{Appendix}. Second, while the discrete-bath HEOM
instabilities in Fig. \ref{Fig. 13} are governed by a single convergence
parameter $L$, the behavior appears quite complicated: note how the
unstable regions merge at lower temperatures and tend to grow more
unstable as $L$ is increased. Since the unstable regions shift as
a function of hierarchy depth, the behavior of the HEOM solution may
be erratic as the hierarchy depth is varied for a fixed temperature,
since instabilities may appear and disappear and vary in severity
at any fixed temperature. In contrast, for continuous-bath HEOM we
see two distinct types of behavior governed by the parameters $K$
and $L$. In Fig. \ref{Fig. 14} we see that as $K$ is incremented,
unstable asymptotes are annihilated one at a time without changing
the temperatures of the remaining asymptotes, effectively lowering
the upper bound on temperatures at which instabilities become problematic.
In Fig. \ref{Fig. 15} we observe that as $L$ increases, the unstable
regions simply change in shape and the temperatures at which asymptotes
occur are invariant. Thus, we find that for continuous-bath HEOM\textbf{
$K$} is predominantly responsible for controlling the most severe
instabilities; an increase in\textbf{ $L$} alone cannot remove the
instability.

Next, we turn to the low-temperature behavior in continuous-bath and
discrete-bath HEOM as it relates to the aforementioned instabilities.
In continuous-bath HEOM, low temperatures are manifestly problematic
since the HEOM are derived using a high-temperature Matsubara expansion.
At $T=0$ continuous-bath HEOM as expressed in Eq. (\ref{eq:-1})
is ill-defined due to the explicit factor of $\cot\left(\beta\gamma/2\right)$
that appears in Eq. \ref{eq:-9}. For small non-zero temperatures
the situation is still problematic; while the HEOM are defined at
temperatures between the asymptotes of $\cot\left(\beta\gamma/2\right)$,
due to the high density of asymptotes (per unit temperature) at low
temperature it is necessary to use a large $K$ to annihilate the
offending asymptotes and obtain converged dynamics. The situation
is quite different in discrete-bath HEOM, where there is no high-temperature
expansion and no notion of Matsubara convergence. It would seem therefore
that discrete-bath HEOM should be amenable to facile low temperatures
simulations as claimed by Chen et al.\citep{Chen2015}. While the
discrete-bath HEOM are indeed well-defined at all temperatures, they
do not eliminate the issue of instabilities at low temperature. Specifically,
the low-temperature instability that in continuous-bath HEOM is controlled
by the number of Matsubara terms reappears in discrete-bath HEOM as
an instability that is controlled by the hierarchy depth. Now that
we have investigated the temperature-dependence of the instabilities,
we focus our attention on methods for overcoming instabilities in
discrete-bath HEOM at $T=0$.
\begin{spacing}{1.2}

\section{Projecting away instabilities exactly\label{sec:Projecting-away-instabilities}}
\end{spacing}

Let us momentarily abandon the goal of extracting a physical $\vec{\rho}(t)$
from the unstable HEOM solution and instead only focus on obtaining
a stable $\vec{\rho}(t)$. This task can be accomplished by projecting
out the unstable modes from the dynamics as follows. Consider the
diagonal matrix
\begin{align}
T(t) & =e^{\Lambda t}.
\end{align}
We construct the projected time evolution matrix $\bar{T}(t)$ by
zeroing out any elements $e^{\lambda_{i}t}$ of $T(t)$ that are larger
than unity in modulus. Then the following projected HEOM solution
will be asymptotically stable,
\begin{align}
\bar{\vec{\rho}}(t) & =V\bar{T}(t)V^{-1}\vec{\rho}(0).
\end{align}
\begin{figure*}
\includegraphics{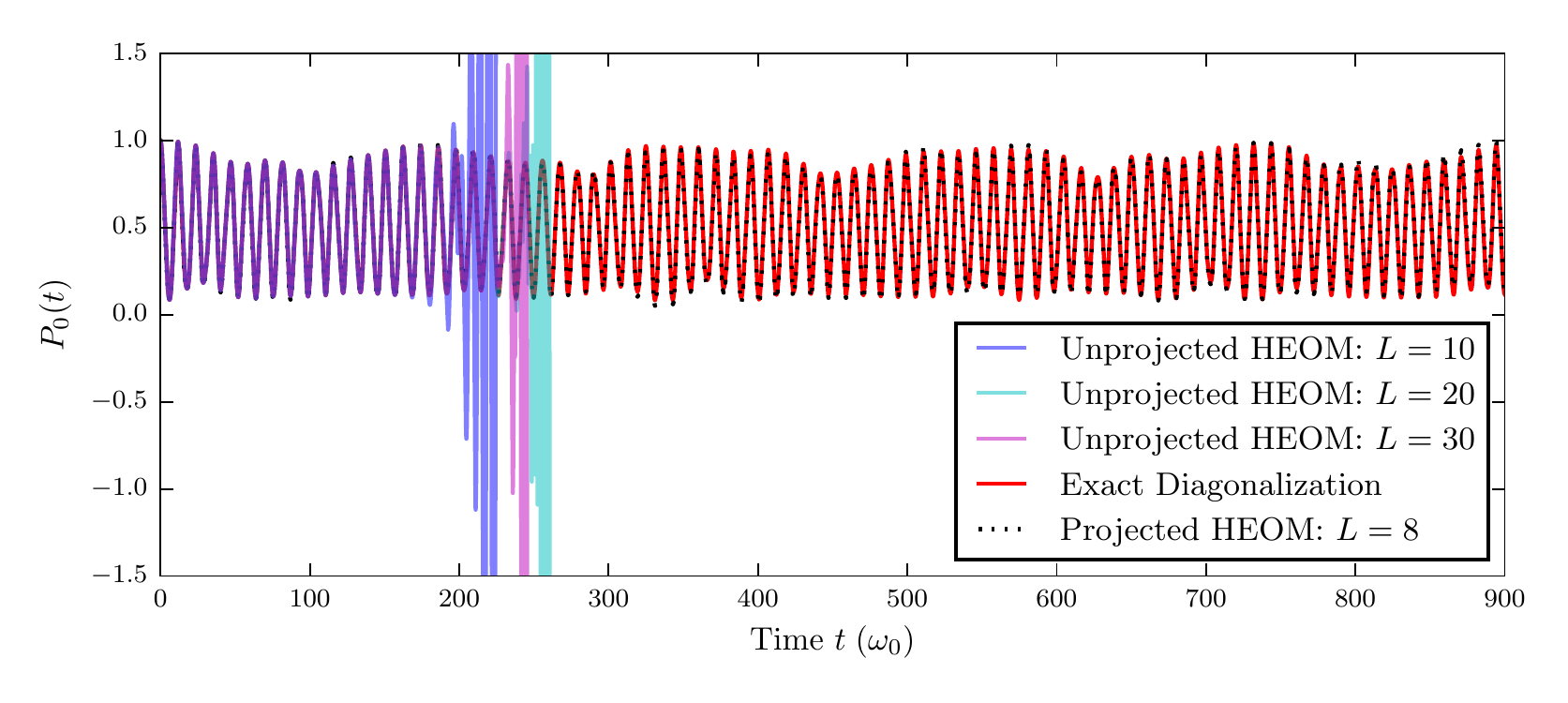}\caption{Electron density on site $0$ for a 3-site Holstein model. First three
curves are computed using HEOM with successively larger hierarchy
depths. Red curve is the exact solution to the Holstein model computed
by diagonalizing the Hamiltonian. Black dotted curve is computed using
HEOM with $L=8$ with unstable modes projected out. Model parameters:
$g=0.3,\omega_{0}=1,J=0.2,T=0$. HEOM calculations are performed using
modified versions of \textit{PHI\citep{Strumpfer2012} }and\textit{
pyrho\citep{Berkelbach}.}}

\label{Fig. 5}
\end{figure*}
Although there is no guarantee that this projection $\vec{\rho}(t)\rightarrow\bar{\vec{\rho}}(t)$
will not unrecognizably alter the dynamics, it turns out that in cases
we have studied for small and intermediate system-bath coupling in
discrete-bath models, $\bar{\vec{\rho}}(t)$ rapidly converges to
the exact $\vec{\rho}(t)$ as the hierarchy depth is increased. This
is unmistakably evident in Fig. \ref{Fig. 5} , where for the 3-site
Holstein model the projection transforms a severely unstable $L=8$
trajectory into what quantitatively resembles the exact dynamics as
computed by diagonalization of the Hamiltonian. Similar success is
observed for the 2-site Holstein model and the DSB. Since the net
effect of unstable modes disappears from the dynamics as $L$ is increased,
we interpret these unstable modes as a spurious, unphysical consequence
of hierarchy truncation. Meanwhile, since the stable modes with $\Re\left[\lambda_{i}\right]\leq0$
approach the exact dynamics as $L$ is increased, we interpret these
stable modes as corresponding to the physical dynamics, which justifies
our projection scheme as a method for obtaining stable, exact dynamics
from discrete-bath HEOM.\\
\begin{spacing}{1.2}

\section{Projecting away instabilities iteratively with Prony filtration\label{sec:Projecting-away-instabilities-1}}
\end{spacing}

\begin{figure*}[t]
\includegraphics{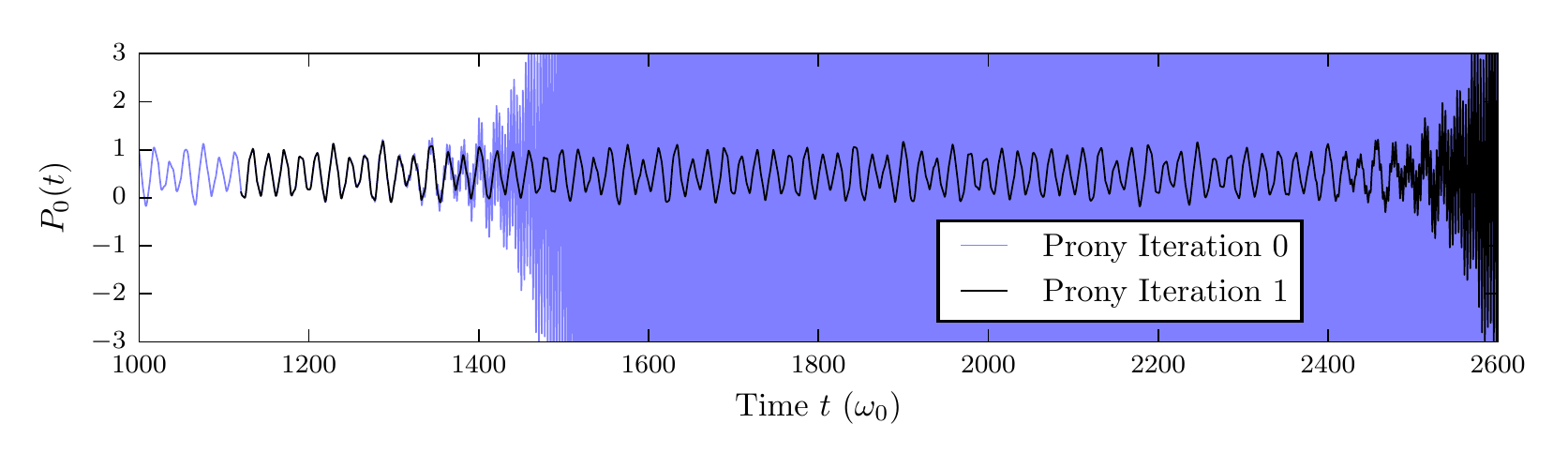}\caption{Single Prony-filtering iteration for the DSB computed with $L=5$.
This plot depicts the electron density on site $0$. First, the blue
curve is computed with HEOM. Next, the instability is approximately
projected out of the blue trajectory, and then the HEOM simulation
is restarted from $t=1120$ at the beginning of the black trajectory
to delay the instability until $t>2400$. Model parameters: $g=0.3,\omega_{0}=1,J=0.2,T=0$.
Calculations are performed using a modified version of \textit{PHI\citep{Strumpfer2012}.}}

\label{Fig. 6}
\end{figure*}

\begin{figure*}
\includegraphics{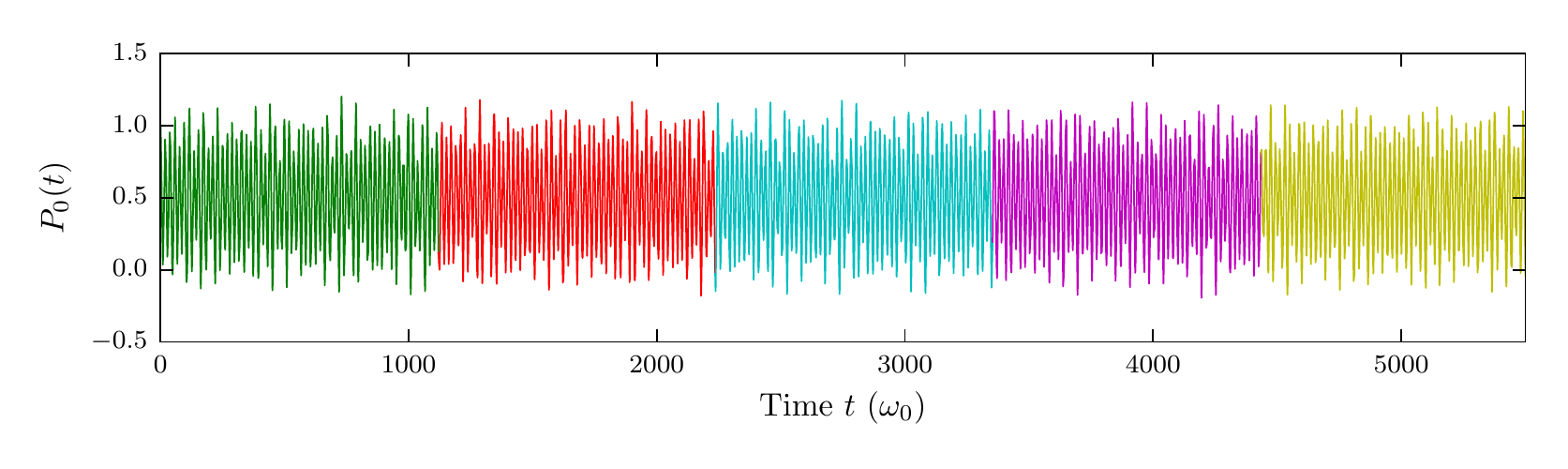}\caption{Extension of the Prony-filtered HEOM simulation from Fig. \ref{Fig. 6}.
Each color represents a different Prony-filtering iteration, with
the instability at times later than $t_{1}$ (see step \ref{enu:Choose-a-time})
not depicted. Model parameters: $g=0.3,\omega_{0}=1,J=0.2,T=0$. Calculations
are performed using a modified version of \textit{PHI\citep{Strumpfer2012}.}}

\label{Fig. 7}
\end{figure*}

\begin{figure}[h]
\includegraphics{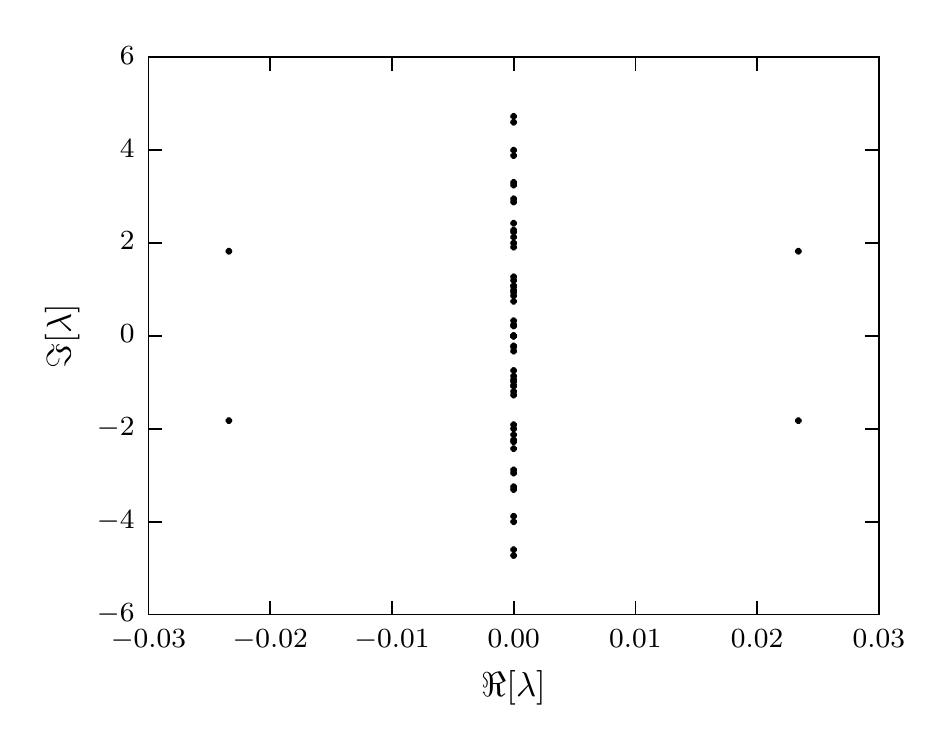}\caption{Spectrum of $A$ for the DSB at $T=0$ with $L=5$. Model parameters:
$g=0.3,\omega_{0}=1,J=0.2$.}

\label{Fig. 12}
\end{figure}

\begin{figure}[h]
\includegraphics{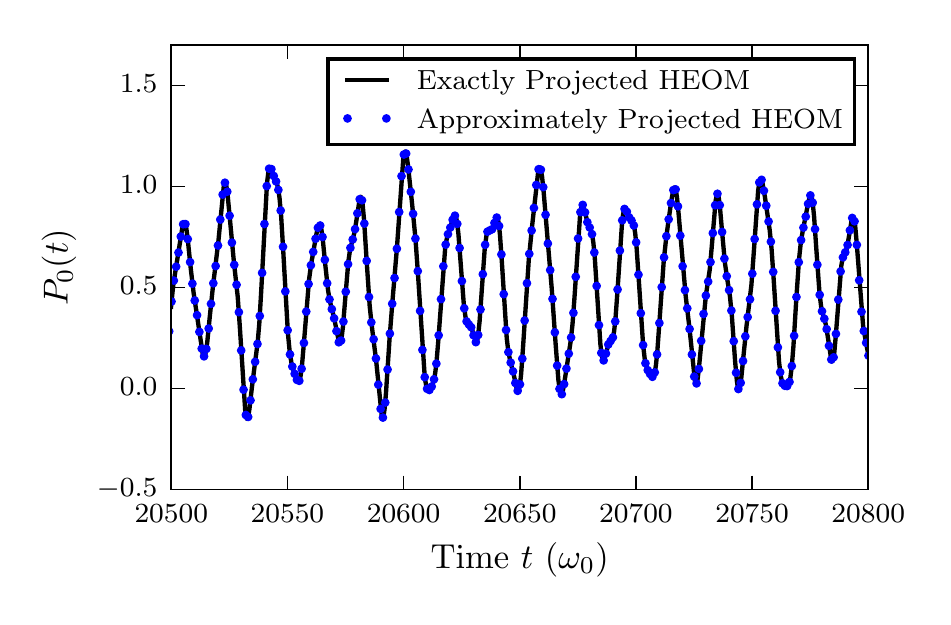}\caption{Projected HEOM trajectories for the DSB at $T=0$ with $L=5$. Black
solid curve uses the diagonalization-based projection. Blue dotted
curve uses the approximate Prony-filtered projection; the data shown
here is from the 19th Prony iteration of the simulation shown in Fig.
\ref{Fig. 7}. Model parameters: $g=0.3,\omega_{0}=1,J=0.2$. Calculations
are performed using modified versions of \textit{PHI\citep{Strumpfer2012}
}and\textit{ pyrho\citep{Berkelbach}.}}

\label{Fig. 8}
\end{figure}

The matrix $A$ in Eq. (\ref{eq:-2}) contains $\tilde{N}^{2}=\left(N^{2}\sum_{n=0}^{L}\frac{(n+2N_{b})!}{n!\left(2N_{b}\right)!}\right)^{2}$
elements\citep{Shi2009} for the discrete-bath case. Therefore, the
$O(\tilde{N}^{3})$ diagonalization-based projection algorithm proposed
in \ref{sec:Projecting-away-instabilities} is completely intractable
for all but the smallest systems, and only then with sufficiently
weak system-bath coupling due to the increased hierarchy depth required
to treat stronger system-bath coupling. As such there is a need for
approximate or iterative computational techniques for removing unstable
modes without requiring an explicit computation of \textbf{all} eigenmodes.
Here we discuss one such algorithm. Since (\ref{eq:-2}) is a first
order linear differential equation, each hierarchy element may be
written as a sum,
\begin{align}
\rho_{i}(t) & =\sum_{j=1}^{\tilde{N}}c_{ij}e^{\lambda_{j}t}\equiv\sum_{j=1}^{\tilde{N}}f_{ij}(t)\equiv\left[\sum_{j=1}^{\tilde{N}}\vec{f}_{j}(t)\right]_{i},\label{eq:-3}
\end{align}
where $c_{ij}$ and $\lambda_{j}$ are complex numbers. The projection
in \ref{sec:Projecting-away-instabilities} is equivalent to subtracting
off all terms in the sum for which $\Re\left[\lambda_{j}\right]>0$.
Consider the following algorithm for subtracting these terms approximately:
\begin{enumerate}
\item Numerically integrate the HEOM using an explicit time-stepping algorithm
such as fourth-order Runge-Kutta until a time $t_{2}$ when $\max_{i}|\rho_{i}(t_{2})|\gg1$.
It is necessary that by $t_{2}$ one or a small number of modes $\{\vec{F}_{j}\}\subset\{\vec{f}_{j}\}$
have grown many orders of magnitude larger than the other modes.
\item \label{enu:For-each-hierarchy}For each hierarchy element, approximate
the weights $c_{ij}=\left(\vec{c}_{j}\right)_{i}$ and complex frequencies
$\lambda_{j}$ of the dominant modes $\{\vec{F}_{j}\}$ using Beylkin
and Monzón's approximate Prony method\citep{Beylkin2005a}. This algorithm
for fitting a function to a sum of complex exponential functions is
described in detail in section 4 of Beylkin and Monzón\citep{Beylkin2005a}.
\item \label{enu:Choose-a-time}Choose a time $t_{1}<t_{2}$ such that $\max_{i,j}|c_{ij}e^{\lambda_{j}t_{1}}|\ll1$.
Subtract the unstable modes from the hierarchy at time $t_{1}$:
\begin{align}
\vec{\rho}(t_{1}) & \rightarrow\vec{\rho}(t_{1})-\sum_{j}\vec{c}_{j}e^{\lambda_{j}t_{1}}.
\end{align}
\item Resume numerical integration from time $t_{1}$ until the next instability
occurs.
\end{enumerate}
In spirit, this algorithm is similar to excited state methods in quantum
mechanics that project out low energy states by imaginary time propagation\citep{Blume1997}.
In Fig. \ref{Fig. 6} we depict a single iteration of the algorithm
above. Iterating this ``Prony-filtering'' algorithm, one can piece
together the same projected HEOM solution that would have been provided
via a single diagonalization using the approach in \ref{sec:Projecting-away-instabilities}.
Such a trajectory is illustrated in Fig. \ref{Fig. 7} for the DSB;
in Fig. \ref{Fig. 12} we show the corresponding spectrum of $A$,
and in Fig. \ref{Fig. 8} we show the essentially perfect agreement
between the diagonalization and Prony-filtering approaches for calculating
stable projected dynamics. 
The Prony filtering approach certainly has a significant prefactor
due to the cost of Runge-Kutta time-stepping and the approximate Prony
analysis of step \enuref{For-each-hierarchy}. However, in a similar
spirit to the power iteration method or Krylov methods for calculating
dominant eigenvectors, this filtration approach has the advantage
of not requiring an $O(\tilde{N}^{3})$ diagonalization to compute
all of the eigenvectors of $A$. Therefore, the filtering algorithm
holds the promise of better scalability compared to the diagonalization-based
projection algorithm.

\section{Nonnormality of HEOM\label{sec:Nonnormality-of-HEOM}}

Many of the matrices typically encountered in quantum mechanics are
Hermitian. Hermitian matrices are diagonalizable and have the properties
that all eigenvalues are real, and that the eigenvectors form an orthonormal
set, i.e. the matrix is normal. We have already seen that the realness
of the eigenvalues is violated for the generator of hierarchy evolution
$A$. It turns out that the latter property is also violated. 

\begin{figure}[t]
\includegraphics{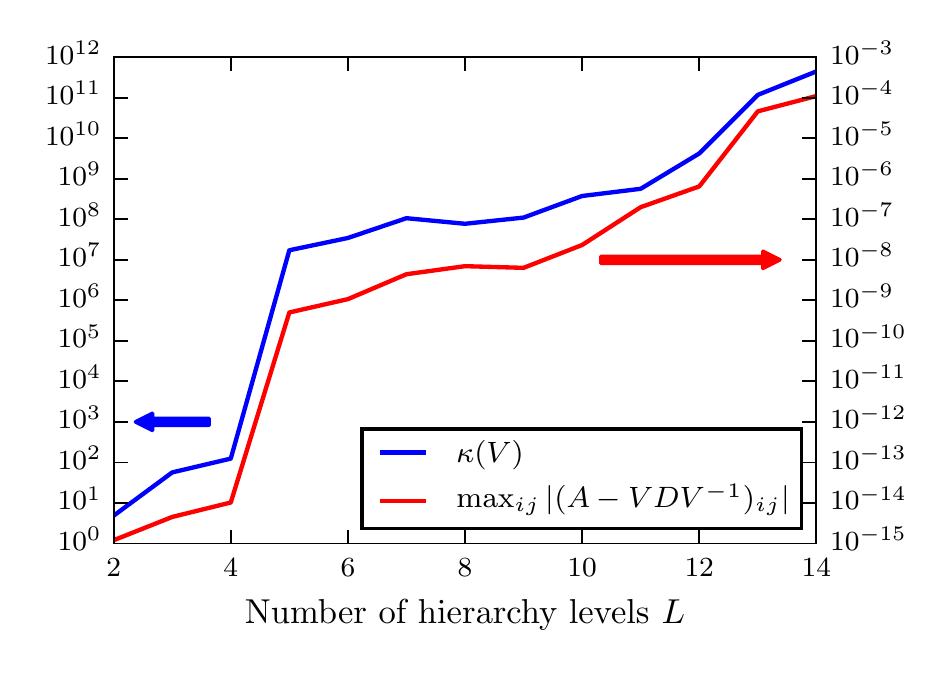}

\caption{Error in computing the eigen-decomposition of $A$ (red, scale on
right axis) and condition number of $V$(blue, scale on left axis)
as a function of hierarchy depth for the 2-site Holstein model at
$T=0$. Model parameters: $g=0.4,\omega_{0}=1,J=0.2$.}

\label{Fig. 10}

\end{figure}

The nonorthogonality of the eigenmodes in HEOM suggests that removal
of instabilities may be ill-conditioned, especially when using deeper
hierarchies as is necessary to treat strong coupling. In Fig. \ref{Fig. 10}
 we show how the numerical error in computing the eigen-decomposition
$A-VDV^{-1}$ grows as a function of hierarchy depth. This numerical
error is not surprising given the nonnormality of $A$, and it explicitly
indicates that the diagonalization-based projection in \ref{sec:Projecting-away-instabilities}
may be not only expensive, but also ill-conditioned. We also show
in Fig. \ref{Fig. 10} how the condition number of $V$ grows with
hierarchy depth. The condition number is defined as 
\begin{align*}
\kappa(V) & \equiv||V||_{2}\times||V||_{2}^{-1},
\end{align*}
where the norm $||V||_{2}$ is the largest singular value of $V$.
This metric exposes how the linear dependence of the eigenmodes of
$A$ grows with hierarchy depth, and it illustrates how their removal
via either diagonalization-based projection or Prony filtration may
be difficult. For calculating the metrics in Fig. \ref{Fig. 10} we
have employed hierarchy scaling\citep{Shi2009} with the aim of reducing
the diagonalization error and condition number. Although the error
is smaller than if we were to use unscaled HEOM, clearly this scaling
does not eliminate the growth with hierarchy depth demonstrated in
Fig. \ref{Fig. 10}.

In addition to the existence of asymptotic instabilities governed
by the spectrum of $A$, it is also possible that numerical instabilities
occur at shorter times due to large transient behavior which is characteristic
of nonnormal dynamics. A classic example of such numerically unstable
transient behavior, as well as its pseudospectral analysis, can be
found in a control theory study of Boeing 767 aircraft\citep{burke2003,Pseudospectra}.
We will leave for future work an investigation of whether the HEOM
instabilities witnessed for larger Holstein models are caused by such
numerically unstable transients or whether they are simply due to
the asymptotic growth of the unstable eigenmodes. It is likely that
hierarchy scaling\citep{Shi2009} substantially reduces such transients
by scaling down the magnitude of elements deep in the hierarchy.

\section{Concluding remarks\label{sec:Concluding-remarks}}

While HEOM is a powerful method that often converges quickly to the
numerically exact dynamics over a significant time range, we have
shown evidence that HEOM trajectories for both continuous-bath and
discrete-bath models at sufficiently low temperature will eventually
hit an exponential wall of instability that completely corrupts the
description of the time evolution. While this instability can typically
be converged away in continuous-bath HEOM, we find that in discrete-bath
HEOM deepening the hierarchy does little to delay instabilities, such
that novel projection schemes are desired. Two methods, direct and
iterative, have been presented to project out the instabilities, and
for discrete-bath HEOM it has been shown that the remaining projected
solution converges to the exact dynamics without requiring many hierarchy
levels. We have discussed challenges that may arise associated with
the computational cost and increasing nonnormality of larger and more
complex HEOM simulations. As of now, we still lack a complete analytical
understanding of the properties of HEOM that lead to the instabilities.
We also fall short of a generic scalable solution for removing these
instabilities. Perhaps with the advent of new HEOM algorithms such
as distributed memory HEOM\citep{Kramer2018} and matrix product state
compressed HEOM\citep{Shi2018}, one may be able to accelerate numerical
integration of the HEOM sufficiently to facilitate projection approaches
such as the Prony filtration approach introduced here. The challenge
of obtaining efficient, stable HEOM solutions will surely benefit
from future work that explores alternative closures to the HEOM which
reduce the instabilities without corrupting the remaining dynamics,
the relation between the breaking of positivity in HEOM\citep{Witt2017}
-- as evidenced by the negative populations in this work -- and
the instabilities, the nature of instabilities in other novel HEOM
formulations\citep{Tang2015,Duan2017,Nakamura2018,Erpenbeck2018},
and computational techniques for removing the unstable modes from
nonnormal, unstable linear systems.

\section*{Supplementary Material}

See \href{http://www.columbia.edu/cu/chemistry/groups/reichman/heomstability.html}{supplementary material}
for animations depicting the spectral dependence of HEOM on hierarchy
depth, number of sites, and choice of model.
\begin{acknowledgments}
The authors thank Gregory Beylkin, Seogjoo Jang, Ramin Khajeh, Benedikt
Kloss, Matthew Reuter, and Qiang Shi for helpful and enlightening
discussions. I.S.D. acknowledges support from the United States Department
of Energy through the Computational Sciences Graduate Fellowship (DOE
CSGF) under grant number: DE-FG02-97ER25308. D.R.R. acknowledges funding
from NSF Grant No. CHE-1839464.
\end{acknowledgments}

\section*{Appendix - Analytical Treatment of CSB\label{Appendix}}

In this appendix we will use a small analytical example to demonstrate
how unstable modes arise in HEOM for the CSB. Consider the HEOM time
evolution operator for the CSB with $K=0,L=2$,\begin{widetext}
\begin{align}
A & =\left[\begin{array}{cc}
-i\mathcal{L} & \Phi\\
\Theta_{0} & -i\mathcal{L}-\gamma
\end{array}\right]=\left[\begin{array}{cccccccc}
0 & -iJ & iJ & 0 & 0 & 0 & 0 & 0\\
-iJ & 0 & 0 & iJ & 0 & 2 & 0 & 0\\
iJ & 0 & 0 & -iJ & 0 & 0 & -2 & 0\\
0 & iJ & -iJ & 0 & 0 & 0 & 0 & 0\\
2i\lambda\gamma & 0 & 0 & 0 & -\gamma & -iJ & iJ & 0\\
0 & -2\lambda\gamma\cot\left(\frac{\beta\gamma}{2}\right) & 0 & 0 & -iJ & -\gamma & 0 & iJ\\
0 & 0 & 2\lambda\gamma\cot\left(\frac{\beta\gamma}{2}\right) & 0 & iJ & 0 & -\gamma & -iJ\\
0 & 0 & 0 & -2i\lambda\gamma & 0 & iJ & -iJ & -\gamma
\end{array}\right].\label{eq:-8}
\end{align}
\end{widetext}\newpage If we consider the case $J=0$, we can get
closed form expressions for the eigenvalues of $A$:
\begin{align}
\left\{ \lambda_{i}\right\} = & \biggr\{0,-\gamma,-\frac{1}{2}\left(\gamma+\sqrt{\gamma^{2}-16\lambda\gamma\cot\left(\frac{\beta\gamma}{2}\right)}\right)\nonumber \\
 & ,-\frac{1}{2}\left(\gamma-\sqrt{\gamma^{2}-16\lambda\gamma\cot\left(\frac{\beta\gamma}{2}\right)}\right)\biggr\}.\label{eq:-10}
\end{align}
Consider further the case where $\lambda>0,\gamma>0$. The last of
these eigenvalues gives rise to an unstable mode whenever
\begin{align}
\Re\left[-\frac{1}{2}\left(\gamma-\sqrt{\gamma^{2}-16\lambda\gamma\cot\left(\frac{\beta\gamma}{2}\right)}\right)\right] & >0.
\end{align}
This condition is equivalent to 
\begin{align}
\cot\left(\frac{\beta\gamma}{2}\right) & <0.
\end{align}
This analytical example reveals alternating temperature regions of
stability and instability for the CSB, akin to those demonstrated
in Figs. \ref{Fig. 14} and \ref{Fig. 15}, with boundaries located
at
\begin{align}
\beta & =\frac{n\pi}{\gamma}\nonumber \\
n & =1,2,...
\end{align}
The asymptotes in the cotangent function at $\beta=2n\pi/\gamma$
correspond to an infinite eigenvalue; at such temperatures the HEOM
specified by Eq. (\ref{eq:-8}) are completely undefined. Furthermore,
for $\beta$ just slightly less than $2n\pi/\gamma$, these HEOM will
be exceptionally unstable due to the large magnitude of the last eigenvalue
in Eq. (\ref{eq:-10}) near the asymptotes of the cotangent function.
On the contrary, for $n\pi<\beta\gamma<\left(n+\frac{1}{2}\right)\pi$
we see that these HEOM are asymptotically stable. These findings agree
with the spectral data shown in Figs. \ref{Fig. 14} and \ref{Fig. 15}.

We should also note that these asymptotes are not present in Fig.
\ref{Fig. 13}, since the HEOM in Eq. (\ref{eq:}) contain $\coth\left(\beta\omega_{0}/2\right)$
rather than $\cot\left(\beta\gamma/2\right)$ and the hyperbolic cotangent
does not contain asymptotes at finite temperatures. Thus, unlike the
behavior exhibited in Figs. \ref{Fig. 14} and \ref{Fig. 15}, that
of Fig. \ref{Fig. 13} is piecewise continuous.

\bibliographystyle{apsrev4-1}
\bibliography{heomstability_v12}

\end{document}